\documentclass[10pts]{article}
\usepackage{graphicx}
\usepackage{bm}

\begin{document}

\title{Radial solitons in armchair carbon nanotubes.}

\author{Alain~Mo\"ise~Dikand\'e \\ 
D\'epartement de Physique, Facult\'e des Sciences, \\
Universit\'e de Douala BP 24157 Douala, Cameroun.} 


\maketitle

\begin{abstract}
Radial solitons are investigated in armchair carbon nanotubes using a 
generalized Lennard-Jones potential. The radial solitons are found in terms of moving kink defects whose velocity 
obeys a dispersion relation. Effects of lattice discreteness on the shape of kink defects are examined by estimating 
the Peierls stress. Results suggest that the typical size for an unpinned kink phase is of the order of a lattice spacing.     
\end{abstract}
                             
\maketitle

\section{Introduction}
\label{sec:level1}
Nanostructured materials are known for their unique properties characterized by rich mechanical, 
structural and electronic properties~\cite{bonard,eklund,kroto} on a short length scale. These particular properties 
give promizing future for various technological applications, and have motivated great recent interest to those materials. 
Among the nanostructured materials recently investigated, carbon nanotubes~\cite{ijima} appear as the most fascinating due 
in part to their cylindrical structure~\cite{eklund,kroto}. Raman spectroscopy~\cite{eklund1,rao} 
has been quite helpful for the understanding of their structural properties, e.g. by allowing direct measurements of their 
vibration spectra. On the one hand, experimental results show that the interplay of the structure and the short length scale gives rise 
to a marked dependence of the frequency of some vibration modes on the tube diameter as for instance the so-called $A_g$ breathing mode. 
On the other hand, combined Raman spectroscopy and inelastic neutron scattering experiments 
reveal pronounced elastic properties as well as plastic deformation processes, two foundamental features 
suggesting that nanotube involve many-body interactions. Owing to their weak 
and dispersive characters, many-body interactions are most often non-local or non-bond forces with van der Waals 
features~\cite{henrard,rao1}. Therefore they can readily be thought of as the main source of stability of nanotubes as 
well as nanotube bundles such as fullerenes. \\
A large amount of work has recently been devoted to the vibration spectra of nanotubes. Being radial, the breathing mode emerges 
as the most influenced by the tube geometry~\cite{rao,rao1,saito}. This feature is manifest in its vibration frequency which 
displays a strong dependence on the tube diameter. Theoretical attempts to explain this dependence was carried out following 
pair-potential approaches and led to a rather simple relationship~\cite{rao1,bandow} agreeing with experiments. 
More explicitely, in these theories many-body potentials are used to deal with the dispersive interactions. The vibrational 
spectrum is thus constructed in terms of the spectrum of phonon excitations in the presence of the dispersive interactions. In 
general, two classes of phonons show up in the Raman spectra namely tangential and radial phonons. \\
The recent interest to excitations in carbon nanotubes was also directed toward modes involving amplitudes more 
larger than phonons. Solitons, which are robust objects describing these large-amplitude excitations, have so far 
been suggested in two main distinct ways~\cite{chamon,menon}. The first~\cite{chamon} relates to the spontaneous lattice 
distorsion responsible for the dimerization of the electronic groundstate of the nanotube. In fact, this first kind of soliton 
excitation is well known in systems undergoing Peierls instabilities, where it forms the state of broken symmetries then 
called domain wall. The second kind of soliton~\cite{menon} follows from the Hamiltonian description of the nanotube dynamics, i.e. 
by writing down a total Hamiltonian consisting of a kinetic energy which describes the tube dynamics, and a potential energy 
accounting for the many-body interaction. The potential energy in this last case connects to the Brenner 
potential~\cite{brenner}, which in the particular context of the paper~ \cite{menon} is expanded for weak interatomic displacements. 
Interestingly, this expansion leads to a generalized nonlinear differential equation which is just the so-called nonlinear Klein-Gordon 
equation, and which can in principle be transformed into any of the known dispersive nonlinear equations as the Nonlinear 
Schr\"odinger equation, Korteweg-de Vrie equation~\cite{remoissenet}, etc. Soliton solutions of these dispersive equations are 
particularly interesting for the dynamics of atomic lattices because of their robustness due to the balance of dispersion by 
nonlinearity both coming from the many-body potential. \\   
In this work, we will attempt to formulate the Hamiltonian description of the soliton excitation in single-wall nanotubes in terms 
of radial solitons. To remain as close as possible to the theory developed in~\cite{menon}, we keep the spirit of 
the many-body interaction generating both the dispersion and nonlinearity in the system dynamics. However, unlike this work we 
will use a generalized Lennard-Jones~\cite{lennard,malino} potential hereafter called "Lennard-Jones hierarchy"(LJh). Let us remark that 
a simple thoeretical model for such an Hamiltonian description follows from the assumption that the carbon-carbon interaction, readily 
regarded as the source of the many-body effects, promotes a $1D$ atomic lattice structure in which atomic sites interact via the many-body 
potential. In single-wall nanotubes, several of these $1D$ atomic lattices result into a nanotube sheet and the effect of some rigid 
background substrate that may provide nonlinearity to the system dynamics as in other contexts~\cite{konto,merwe} is rather unlike. In fact, 
being a nonlinear function of the lattice displacement, the many-body potential will furnish the necessary dispersion and nonlinearity needed 
to stabilize the lattice excitations. \\
Competing effects between dispersion, nonlinearity and lattice discreteness can become quite relevant to the existence and stability of some 
of these nonlinear excitations. For instance, kink-soliton defects will show a rather strong sensibility vis-\`a-vis 
the discrete lattice structure and will most often need a threshold width to survive lattice discreteness effects. To this viewpoint, it is 
well established that narrow kinks are more prone to form pinned soliton states~\cite{dika2}, proving that pinning effects can have 
dramatic incidence on transport properties of materials~\cite{pok,bak1,bak2}. As we will be interested with kink-soliton defects, it will be 
instructive to look at the effects of the comptetition between the two main ingredients of the model i.e. the dispersion and nonlinear, on the 
continuum shape of a kink-soliton defect propagating on the discrete $1D$ lattices forming the single-wall nanotube.   \\
To start, we investigate the dynamics of radial kink-soliton defects in the armchair nanotube for many-body potentials belonging to the LJh. 
We will also examine dispersion features of the continuum kink-soliton solution in the $1D$ discrete lattice and derive a dispersion law, valid for 
a given $(m,m)$ armchair configuration. Next, the effects of lattice discreteness on shapes of the radial kink-soliton defects will be 
considered. In this context, the analytical expression of the Peierls stress~\cite{nabarro,willis,dika1} will be derived. 
\section{The model, soliton solutions and dispersions.}
\label{sec:level2}
\begin{figure}
\centering
\includegraphics{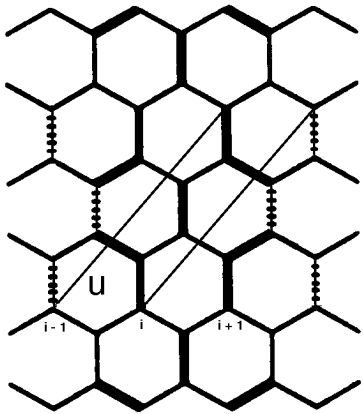}
\caption{Sketch of the one-dimensional lattice of atoms in an armchair nanotube. {\bf{u}} is a unit vector.}
\label{scheme}
\end{figure}

\begin{figure}
\includegraphics{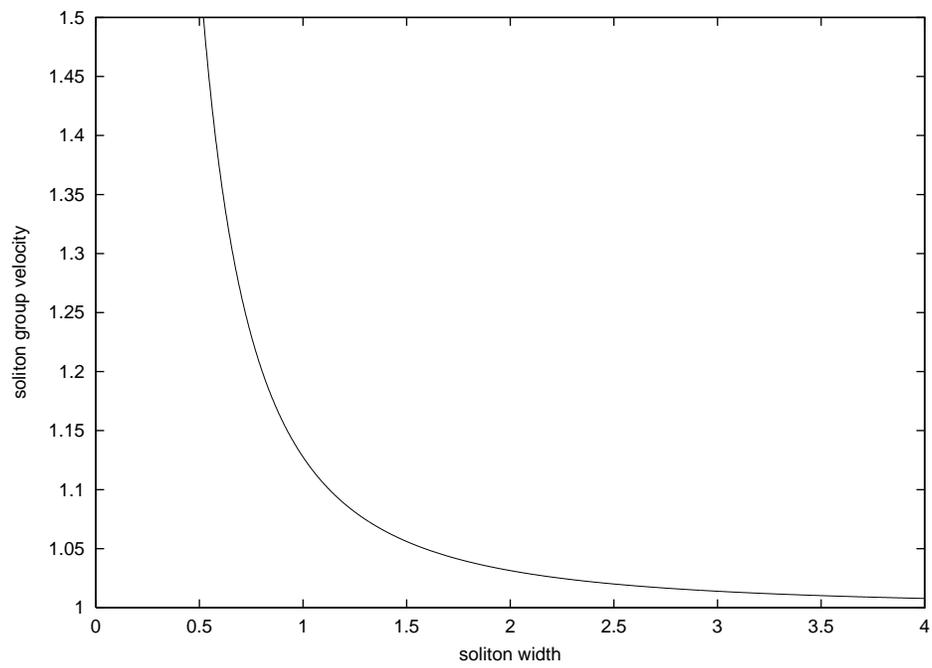}
\caption{The reduced group velocity of the kink defect as a function of the dimensionless soliton width(i.e. in unit of the lattice constant $a_o$).}
\label{vit}
\end{figure}

Consider a $1D$ lattice consisting of an atomic backbone in which neihgbouring atoms interact via 
a pair potential of one of the LJhs $(n_1,n_2)$. Denoting by $r_n$ the relative displacement of the 
$n^{th}$ atom, the total energy of the lattice can be written:
\begin{equation}
E= \sum_{n=1}\left[\frac{M}{2} \dot{r}_n^2 +  V_{LJh}(r_{n+1} - r_n)\right] \label{a1}
\end{equation}
where dot refers to derivative with respect to time $t$, $M$ is the total mass associate to the $n^{th}$ site and $V_{LJh}$ is the LJh potential 
assumed of a general form:
\begin{equation}
V_{LJh}(r)= 4\epsilon_o\left[\left(\frac{\sigma}{r}\right)^{n_1} - 2 \left(\frac{\sigma}{r}\right)^{n_2}\right]  \label{a2}
\end{equation}
$n_1$ and $n_2$ in this last formula are integers chosen such that $n_1 > n_2$, while $\sigma$ is a characteristic 
length and $\epsilon_o$ a characteristic energy. The LJh potential~(\ref{a2}) possesses two different 
interaction branches i.e. a short-range repulsive branch and a long-range(van der Waals-like) attractive 
branch. The two branches meet at the equilibrium point: 
\begin{eqnarray}
r_o= \left(\frac{n_1}{2n_2}\right)^{\frac{1}{n_1-n_2}}\sigma, \hspace{.1in} V_{LJh}(r_o)= -4\left(\frac{n_1}{2n_2}\right)^{\frac{n_2}{n_2-n_1}}\epsilon_o  \label{a3}
\end{eqnarray}
corresponding to the dissociation point of the lattice. For the bare LJ potential, $(n_2,n_1)=(6,12)$ suggesting a 
potential minimum at $r_o= \sigma$, for which $V_{eq}= -4\epsilon_o$. Another frequent set of values~\cite{malino} 
is $(n_2, n_1)=(3,9)$ for which $r_o= \left(\frac{3}{2}\right)^{1/6} \sigma$ and 
$V_{eq}= -4\left(\frac{2}{3}\right)^{1/2}\epsilon_o$. \\
 The characteristic distance $\sigma$ will generally depend on 
geometry of the material. Nevertheless, by appropriate means we can also define a square frame in an hexagonal geometry 
and parametrize the 
displacement vectors ${\bf{r}_n}$ by defining radial variables $\theta_n$ with respect to the new frame, obtaining:
\begin{equation}
{\bf{r}}_{n+1} - {\bf{r}}_n= a_{n+1} \hat{u}_{n+1} - a_i \hat{u}_n, \label{relatif}
\end{equation}
where $\hat{u}_n$ are unit vectors along rods lying on the sheet of nanotube as in figure~\ref{scheme}, while resting on 
atoms(labeled $i$, $i+1$, ...) of the substrate backbone. 
At equilibrium, the rods align uniformly with an equilibrium angle 
$\theta_o$ with respect to the $1D$ lattice. The equilibrium angle $\theta_o$ can thus be assumed as the chiral angle of 
the nanotube, and for an armchair configuration $\theta_o$ will be determined by the "chiral coordinate" $(m, m)$ which 
also fixes the diameter of the tube. Therefore, we readily expect this last quantity to affect excitations of the tube. A 
simple way to implement this dependence is to consider instead the twist motion of the entire rod then assumed rigid. In this way, the mass M turns to 
an effective mass resting on a reference atom on the substrate lattice. Expressing $\hat{u}_n$ as a function of the radial 
coordinates and discarding the constant chiral angle, we obtain:
\begin{equation}
\hat{u}_n= \hat{e}_x\cos \theta_n + \hat{e}_y \sin \theta_n. \label{a5}
\end{equation}
Since we assume rigid rods, we can set $a_{n+1}=a_n=\ell_o$. Furthermore, we place ourselves on an orthonormal basis where $\hat{e}_{i=x, y}$ are orthonormalized unit vectors. 
From ~(\ref{a5}) and taking into account all the considerations made above, the separation between the $(n+1)^{th}$ and the $n^{th}$ site 
block along the $1D$ lattice reduces to:
\begin{equation}
 r^2= 2\ell_o^2 \left[1 - \cos \left( \theta_{n+1} - \theta_{n} \right) \right] \label{a6}
\end{equation}   
We will now proceed to the key transformation of our approach, consisting of an expansion of~(\ref{a2}) in Taylor series 
arriving at:
\begin{equation}
V_{LJh}= V_o(n_1,n_2) + \sum_{k=1}^{\infty}V_k(n_1,n_2) \cos^k \left(\theta_{n+1} - \theta_n\right) \label{a7}
\end{equation}
where we have set:
\begin{equation}
V_k(n_1,n_2)= 4 \left[b_k^{(n_1)} \left(\frac{\sigma}{\sqrt{2}\ell_o}\right)^{n_1} - 2 b_k^{(n_2)} \left(\frac{\sigma}{\sqrt{2}\ell_o}\right)^{n_2} \right]\epsilon_o \label{a71}
\end{equation}
\begin{equation}
b_k^{(j)}= \frac{(-1)^k}{k!} \frac{j}{2} \frac{(j+2)}{2} ... \frac{(j + 2k - 2)}{2}  \label{deve}
\end{equation}
This expansion is a simple generalization of the Taylor series representation of the LJ potential, except that the 
condition of weak relative displacements $r_{n+1} - r_n$ is replaced by its equivalent involving the relative angle 
$\theta_{n+1} - \theta_n$. Similar expansion has also been used in the study of critical properties of lipid monolayers~\cite{kreer}. \\
When $k=1$, equations~(\ref{a7})-~(\ref{a8}) lead to a one-period sinusoidal potential 
i.e. the sine-lattice(SL) potential~\cite{sine1,sine2,sine3,sine4,sine5}. To complete our transformation, we also need to 
formulate the kinetic energy in terms of the discrete radial variables. This is achieved by introducing the 
elementary displacement $d\, r_n \sim \ell_o d\theta_n$. With this assumption, the total Hamiltonian~(\ref{a1}) reduces to the 
SL Hamiltonian i.e.: 
\begin{eqnarray}
E_{SL}= \sum_{n=1}\left[\frac{M \ell_o^2}{2} \dot{\theta}_n^2 +  V_{SL}(\theta_{n+1} - \theta_n)\right], \nonumber \\
V_{SL}(\theta_{n+1} - \theta_n)\}= V_1(n_1, n_2) \, cos(\theta_{n+1} - \theta_n) \label{slat}
\end{eqnarray}
from which we derive the following set of difference differential equations:
\begin{equation}
\theta_{n, tt} - \omega_o^2\left[\sin \left( \theta_{n+1} - \theta_n \right) + \sin \left( \theta_{n-1} - \theta_n \right) \right]= 0 \label{a8}
\end{equation}
with:
\begin{equation}
\omega_o^2= -\frac{V_1(n_1,n_2)}{M \ell_o^2} \label{a9}
\end{equation}
Equation~(\ref{a9}), also called SL equation~\cite{sine1,sine2,sine3,sine4,sine5}, was first introduced for conformational transitions 
in DNA macromolecules. It admits, at large radial displacements, a kink-soliton solution described by the ansatz:
\begin{equation}
\theta_n(t)= 2 \arctan \exp \left(q\, n\, a_o - \omega t\right) \label{a10}
\end{equation} 
where $a_o$ is the equilibrium separation between two neighboring atomic sites along the $1D$ lattice. The 
ansatz~(\ref{a10}) can also be looked on as a kink-soliton defect that propagates along the $1D$ lattice 
being subjected to the dispersion law: 
\begin{equation}         
\omega^2= 4\omega_o^2\sinh^2(q a_o/{2})   \label{a11}
\end{equation} 
In this last relation, the quantity $\omega_o$ acquires more physical meaning in terms of the characteristic frequency 
of kink vibrations on the $1D$ lattice. According to the argument of~(\ref{a10}), the characteristic width of the 
kink defect is inversely proportional to $q$ while the kink velocity is determined by the 
dispersion relation~(\ref{a11}). To this viewpoint, from~(\ref{a11}) a group velocity can indeed be defined as follow:
\begin{equation} 
\vartheta(q) = \frac{\partial \omega}{\partial q} , \label{vites}
\end{equation}
This group velocity displays a threshold value in the limit $q \rightarrow 0$ and is given by $\vartheta_o= \omega_o a_o$. 
Returning to formula~(\ref{a9}), we will notice that ignoring the constant $V_1(n_1,n_2)$ 
which depends mainly on details of the many-body potential, the characteristic frequency $\omega_o$ appears as an inverse function 
of the tube diameter $\ell_o$. It turns out that the threshold group velocity $\vartheta_o$ will dependent on the ratio between the 
two characteristic length scales of the nanotube i.e. the lattice constant $a_o$ and the tube diameter $\ell_o$. While $a_o$ can be 
assumed a free parameter, $\ell_o$ is a function of the chiral coordinate $(m,m)$ meaning that the ratio $a_o/ \ell_o$ 
will acquire distinct values for different armchair configurations. \\ 
On figure~\ref{vit}, we plot the reduced group velocity $\vartheta(q)/\vartheta_o$ as a function of the kink width 
$1/q$. We observe that the kink velocity gets slowed down when its width increases. Moreover, the group velocity rapidly 
saturates to the minimum threshold $\vartheta_o$ at finite, but relatively large values of the kink width. 
It is useful to underline the qualitative agreement between the behaviour reflected by curve of figure~\ref{vit} 
and result of ref.~\cite{menon} on longitudinal solitons(see figure 2 in this paper) which we recall it, uses the Brenner potential. \\
We will close this section by looking at the energetics of the radial soliton defect obtained in~(\ref{a10}). In particular 
we estimate the defect energy which in the usual language, is the characteristic energy needed to create a kink 
soliton of this form. Replacing~(\ref{a10}) in the continuum version of the SL 
Hamiltonian~(\ref{slat})(i.e. making the continuum limit approximation $n a_o \rightarrow x$) and taking into account the dispersion relation~(\ref{a11}), we find:
\begin{equation}
E= \frac{M \ell_o^2 \omega^2}{q a_o}  \label{a12}
\end{equation}
This quantity is vanishing at large defect widths, a behaviour already displayed by the kink frequency or also the dispersion 
relation~(\ref{a11}). 
 \section{Discrete lattice effects on the kink-soliton defects.}
 \label{level3}
\begin{figure}
\includegraphics{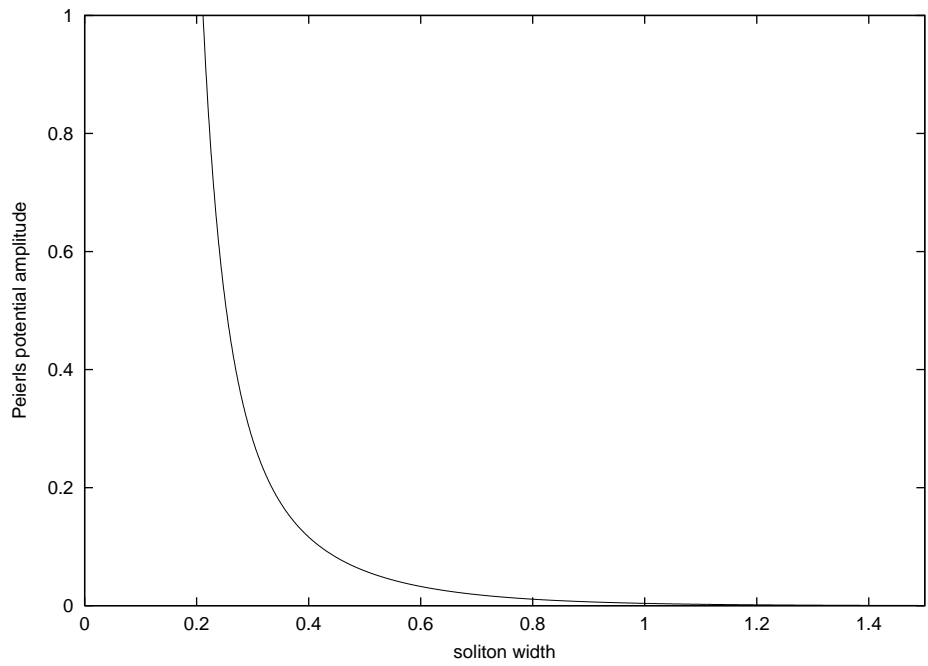}
\caption{Amplitude of the Peierls stress, plotted versus the dimensionless soliton width(i.e. in unit of the lattice constant $a_o$).}
\label{fig:peie}
\end{figure}

The analytical expression~(\ref{a10}) is actually not an exact solution of the nonlinear equation~(\ref{a8}), 
but describes a continuum soliton solution accomodating the $1D$ discrete lattice. Then, in addition to the dispersion relation~(\ref{a11}) 
which dictates the propagation velocity of the continuum kink soliton, the discreteness of the propagation medium will introduce an 
additional constraint for the stability of shape of this continuum kink soliton in the $1D$ discrete lattice. This constraint acts as a mechanical stress that tends to 
keep the soliton pinned to the discrete background of the lattice, hence it is comparable to the Peierls stress commonly observed 
in crystal monolayer as well as multilayer phenomena~\cite{lothe}. The concept of second Peierls stress is best known 
for systems involving a rigid background substrate, such as in the Frenkel-Kontorova~\cite{konto} and Frank-van der Merwe~\cite{merwe} models, 
in addition to the strain energy of a Hookes type. Nevertheless, in a recent 
paper~\cite{dika1} we suggested an approach to deal with the Peierls-Nabarro effect for model Hamiltonians of the SL type and will 
follow this work. Starting, remark that taken in its current analytical form i.e. keeping only the site coordinate $n$, 
the continuum kink soliton~(\ref{a10}) looks strongly dependent on the discreteness and will not survive pinning to the lattice sites. 
To avoid this dramatic lattice discreteness effect, we shift the kink position by introducing 
a center-of-mass coordinate $n_o$ that preserves both the kink translational-invariant shape and dispersion governed by 
the dispersion relation(\ref{a11}). With these considerations the kink shape now becomes:
\begin{equation}
\theta_n(t)= 2 \arctan \exp \left[q\, (n - n_o) a_o\right] \label{solit}
\end{equation} 
Inserting~(\ref{solit}) in the strain term of the Hamiltonian~(\ref{slat}) and Fourier transforming as in~\cite{dika1}, 
we obtain: 
\begin{equation}
\epsilon_p(q)= \frac{2 \pi^2 M \ell_o^2 \omega^2}{a_o q^2 sinh(2 \pi^2/q a_o)}      \label{pnab}
\end{equation}
$\epsilon_p$, or the Peierls-Nabarro potential barrier for the SL model~(\ref{slat}), is plotted on figure~\ref{fig:peie} versus the soliton width.
Here we use the quantity: 
\begin{equation}
V_o= 8 \pi^2 Ma_o \ell_o^2 \omega_o^2. \label{constant}
\end{equation}
as unit of mechanical stress. The physical meaning of the decrease of the Peierls-nabarro potential barrier with an increase of the soliton width is 
a reduced effect of the mechanical stress for relatively large soliton defects. Typically, a soliton defect of size about one or a few lattice spacings will 
almost be insensitive to the lattice discreteness as suggested by the drastic fall-off in figure~\ref{fig:peie}. What this result implies as for the physics of 
the nanotube is the possibility of a conducting state dominated by narrow soliton defects. 
\section{\label{sec:level3}Summary and concluding remarks}
We have explored an Hamiltonian description for radial soliton excitations and dispersions in single-walled nanotubes with an armchair 
configuration 
using a generalized Lennard-Jones potential. A first point to underline is the agreement between our results and a previous study~\cite{menon} on longitudinal 
solitons using Brenner's potential. This includes the velocity and energy whose variations with the soliton width are 
qualitatively similar to predictions of this previous work. In addition, we found that dispersion confines the soliton velocity within a 
defined interval characterized by a minimum threshold velocity proportional to the ratio of the lattice spacing to the tube diameter. \\
Effects of the lattice discreteness on the soliton shape have also been investigated. We obtained that the threshold width for an unpinned 
soliton phase was nearly of the order of a lattice spacing. This narrow soliton-induced unpinned kink defect phase is the 
signature of a dominant dispersion and may therefore well be understood as reflecting the absence of a single-particle potential. 
Indeed, in general the single-particle potential will enhance the threshold width for the unpinned kink phase~\cite{willis} owing to the rigid 
character of the substrate to which it relates. Whether this single-particle substrate potential is pertinent to single-wall nanotubes and/or 
their derivatives i.e. nanotube bundles, fullerenes, etc., is to our view a central issue for a more realistic description of the nonlinear 
dynamics of these systems.

\end{document}